\documentclass[]{aa}
\usepackage[dvips]{graphicx}

\begin{document}

\title {Effects of color superconductivity on the nucleation of quark matter in neutron stars}

\author{I. Bombaci \inst{1}, G. Lugones \inst{1} \and  I. Vida\~na \inst{2}}
\institute{Dipartimento di Fisica ``Enrico Fermi'' Universit\`a di
Pisa and INFN Sezione di Pisa, \\ Largo Bruno Pontecorvo 3, 56127
Pisa, Italy \\ \email{bombaci@df.unipi.it} ;
\email{lugones@df.unipi.it} \and Departament d'Estructura i
Constituents de la Materia, Universitat de Barcelona,\\ Avda.
Diagonal 647, E-08028 Barcelona, Spain \\ \email{vidana@ecm.ub.es}}

\abstract{}{We study the nucleation of quark matter drops at the
center of cold deleptonized neutron stars. These drops can be made
up by unpaired quark matter or by color superconducting quark
matter, depending on the details of the equations of state for quark
and hadronic matter.  The nature of the nucleated phase is relevant
in the determination of the critical mass $M_{cr}$ of hadronic stars
above which it is possible a transition to a quark star (strange or
hybrid).}
{We investigate the dependence of  $M_{cr}$ upon the parameters of
the quark model (the Bag constant $B$, the pairing gap $\Delta$, and
the surface tension $\sigma$ of the quark-hadron interface) and for
different parametrizations of the hadronic equation of state. We
also calculate the energy released in the conversion of a pure
hadronic star having the critical mass into a quark star.}
{In general, the dependence of $M_{cr}$ on  $B$, $\Delta$ and
$\sigma$ is mild if the parameters of the quark model correspond to
hybrid stars, and strong if they correspond to strange stars. Also,
the critical mass always decreases with $\Delta$, and increases with
$B$ and $\sigma$.  The total released energy is in the range $3
\times 10^{52} \mathrm{erg}$ - $4 \times 10^{53} \mathrm{erg}$.}
{For a large part of the parameter space corresponding to hybrid
stars,   the critical mass is very close (but smaller than) the
maximum mass of hadronic stars, and therefore compatible with a
``mixed'' population of compact stars (pure hadronic up to the
critical mass and hybrid above the critical mass). For very large
$B$ the critical mass is never smaller than the maximum mass of
hadronic stars, implying that quark stars cannot form through the
here studied mechanism. The energy released in the conversion is
sufficient to power a gamma ray burst. }

\keywords{dense matter, elementary particles, stars: neutron}

\date{Received March 2006}

\titlerunning{Effects of color superconductivity... }
\maketitle

\section{Introduction}

The nucleation of quark matter in neutron stars has been studied by
many authors, due to the potential connection with explosive
astrophysical events such as supernovae and gamma ray bursts. Some
of the earlier studies on quark matter nucleation (see e.g. Horvath
et al. 1992, Horvath 1994, Olesen \& Madsen 1994, and references
therein) dealt with thermal nucleation in hot and dense hadronic
matter. In these studies, it was found that the prompt formation of
a critical size drop of quark matter via thermal activation is
possible above a temperature of about 2 -- 3 MeV. As a consequence,
it was inferred that pure hadronic stars are converted to strange
stars or to hybrid stars within the first seconds after their birth.
However, neutrino trapping in the protoneutron star phase strongly
precludes the formation of a quark phase (Lugones and Benvenuto
1998, Benvenuto and Lugones 1999, Vida\~na, Bombaci \& Parenti
2005). Then, it is possible that the compact star survives the early
stages of its evolution as a pure hadronic star. In this case,
nucleation of quark matter would be triggered by quantum
fluctuations  in degenerate (T = 0) neutrino-free hadronic matter
(see e.g. Grassi 1998,  Iida and Sato 1998, Berezhiani et al. 2003,
Harko, Cheng and Tang 2004, Bombaci, Parenti and Vida\~na 2004, and
references therein).

A general feature of degenerate Fermi systems is that they become
unstable if there exist any attractive interaction at the Fermi
surface (Bardeen, Cooper and Schrieffer 1957). This instability
leads to the formation of a condensate of Cooper pairs and the
appearance of superconductivity. In QCD any attractive quark-quark
interaction will lead to pairing and color superconductivity
(Barrois 1977, Bailin and Love 1984 and references therein). Since
the typical superconducting gaps in quark matter may be as large as
$\Delta \sim 100$ MeV (see e.g. Alford 2001 and references therein)
it is interesting to study the effects of color superconductivity in
the process of nucleation. Some recent studies have gone in this
direction (Drago, Lavagno \& Pagliara 2004, Lugones and Bombaci
2005), but we shall give here a more complete and self-consistent
study.

In a recent work (Lugones \& Bombaci 2005, hereafter LB05) we have
studied the formation of superconducting quark matter in bulk,
paying particular attention to the microscopic state of quark matter
just after the deconfinement. As explained in LB05, several states
are possible in principle. For example, quantum fluctuations could
form a drop of $\beta$-stable quark matter (hereafter the
$Q^{\beta}$ phase). However, this is strongly suppressed with
respect to the formation of a non $\beta$-stable drop by a factor
$\sim G_{\mathrm{Fermi}}^{2N / 3}$ being $N$ the number of particles
in the critical size quark drop. This is so because the formation of
a $\beta$-stable drop would involve the almost simultaneous
conversion of $\sim N/3$ {up and down} quarks into strange quarks.
For a critical size $\beta$-stable nugget at the center of a neutron
star it is found $N \sim 100-1000$, and therefore the factor is
actually tiny. This is the same reason that impedes that an iron
nucleus converts into a drop of {strange} quark matter, even in the
case in which strange quark matter had a lower energy per baryon
(Bodmer-Witten-Terazawa hypothesis).

However, quantum fluctuations can form the so called Q*-phase
bubbles, in which the flavor content of the quark phase is equal to
that of the $\beta$-stable hadronic system at the same pressure.
Since no flavor conversion is involved, there are no suppressing
Fermi factors, and a Q*-phase drop can be nucleated much easier.
Once a critical size drop of the Q*-phase is formed the weak
interactions will have enough time to act, changing the quark flavor
fraction of the deconfined droplet to lower its energy, and a
droplet of the ($\beta$-stable) $Q^{\beta}$-phase is formed. Notice
that in degenerate matter, the intermediate phase can be made up by
unpaired quark matter (hereafter the $Q^*_{unp}$ phase) or by
color-superconducting quark matter ($Q^*_{\Delta}$ phase). Which one
of the two kind of droplets ($Q^*_{unp}$ or $Q^*_{\Delta}$) will
nucleate depends on the Gibbs free energy per baryon of each phase
($g_{unp}$, $g_{\Delta}$), as discussed in detail in LB05.

The analysis in LB05 has been made in bulk, i.e. without taking into
account the energy cost due to finite size effects in creating a
drop of deconfined quark matter in the hadronic environment. As a
consequence of the surface effects it is necessary to have an
overpressure $\Delta P = P - P_0$ with respect to the bulk
transition point $P_0$ (see Fig. 1 in LB05). Thus, above $P_0$,
hadronic matter is in a metastable state, and the formation of a
real drop of quark matter occurs via a quantum nucleation mechanism.
In this work we shall study the quantum nucleation process including
the effect of color superconductivity, and analyze the implications
for neutron stars. \footnote{It is worth mentioning that Drago,
Lavagno \& Pagliara 2004 have investigated the nucleation of a
superconducting $\beta$-stable phase made up by color-flavor-locked
(CFL) quark matter. However, as stated before, the direct nucleation
of a drop of such a phase is disfavored by weak interactions.}

A final comment is worthwhile concerning the formation of mixed
hadron-quark phases, in which the electric charge is zero
\textit{globally} but not \textit{locally}, i.e. the two phases have
opposite charge (Glendenning 1992, 2001).  As discussed in Lugones
and Benvenuto 1998, mixed phases cannot form in the here studied
\emph{just-deconfined} phase. The reason is that the flavor
conservation condition guarantees that a just-deconfined
quark-matter drop has \textit{initially} exactly the same electric
charge than the hadronic drop from which it originated (i.e. zero).
Of course, charge separation could occur later on (if energetically
preferred) and a mixed phase could form. However, notice that Debye
screening effects and the surface tension can prevent mixed phases
to form (see e.g. Tatsumi \& Voskresensky 2003, Endo et al. 2005,
Maruyama et al. 2006, and references therein). In any case, the
study of mixed phases concerns the state of the system at times much
larger than the ones that are addressed in this paper.

\section{The properties of the intermediate non-$\beta$-stable
quark phase}

As emphasized in the Introduction  the intermediate phase can be in
an unpaired state ($Q^*_{unp}$) or in the so-called two-flavor color
superconducting  (2SC) state  (hereafter $Q^*_{\Delta}$). In this
section we resume the equation of state of the $Q^*_{\Delta}$-phase
(for more details see LB05).

The physical conditions imposed on the $Q^*_{\Delta}$-phase are
flavor conservation and color neutrality of the quark gas. Flavor
conservation means that the particle number (per baryon) of quarks
$u$, $d$ and $s$ are the same in the hadronic phase and in the
$Q^*_{\Delta}$-phase. The only difference is that in the hadronic
phase quarks are confined inside hadrons, and in the
$Q^*_{\Delta}$-phase they are deconfined and paired. This condition
can be expressed  in terms of two parameters $\xi$ and $\eta$:
\begin{equation}
n_d = \xi ~ n_u. \label{h1}
\end{equation}
\begin{equation}
n_s = \eta ~ n_u. \label{h2}
\end{equation}
\noindent where $n_i$ is the particle number density of the
$i$-species in the quark phase. The quantities $\xi \equiv Y^H_d /
Y^H_u$ and $\eta \equiv Y^H_s / Y^H_u$ are functions of the
pressure, and characterize the composition of the hadronic phase.
These expressions are valid for \textit{any} hadronic EOS. For
hadronic matter containing $n$, $p$, $\Lambda$, $\Sigma^{+}$,
$\Sigma^{0}$, $\Sigma^{-}$, $\Xi^{-}$, and $\Xi^{0}$, we have
\begin{eqnarray}
\xi &=& \frac{n_p  +  2  n_n  + n_{\Lambda} + n_{\Sigma^{0}} +  2
n_{\Sigma^{-}}  + n_{\Xi^{-}}}{2  n_p  +  n_n  +  n_{\Lambda} + 2
n_{\Sigma^{+}} + n_{\Sigma^{0}}  +  n_{\Xi^{0}}}, \\
\eta &=& \frac{n_{\Lambda}  + n_{\Sigma^{+}} + n_{\Sigma^{0}}  +
n_{\Sigma^{-}} + 2 n_{\Xi^{0}} + 2 n_{\Xi^{-}}}{2  n_p  +  n_n  +
n_{\Lambda} + 2 n_{\Sigma^{+}} + n_{\Sigma^{0}}  +  n_{\Xi^{0}}}.
\end{eqnarray}
Notice that $\xi$ and $\eta$ determine univocally the number of
electrons present in the system through electric charge neutrality
of the deconfined phase:
\begin{equation}
3 n_{e} = 2 n_{u} - n_{d} - n_{s} .\label{h3}
\end{equation}

We also impose two pairing conditions:  $n_{dr}= n_{ug}$ in order to
allow for paring between quarks $d_r$ with $u_g$, and  $n_{ur}=
n_{dg}$ in order to allow for paring between quarks $u_r$ with
$d_g$. Finally, the system must be globally colorless, that is:
$n_r = n_g = n_b$.

From the above equations we obtain the number densities of each
quark species in the paired phase as functions of the flavor
composition $\eta$, $\xi$:
\begin{eqnarray}
{n_{ub}} & =& \frac{ 4 - 2\xi}{1  +  \xi} ~ {n_{ur}} \label{n1}\\
{n_{db}} & = & \frac{-2  + 4 \xi}{1 + \xi} ~{n_{ur}} \label{n2}\\
{n_{sb}} & = & \frac{2 \eta }{1  + \xi} ~{n_{ur}} \label{n3}\\
{n_{e}} & = & \frac{2 (2 - \eta - \xi )}{1 + \xi} ~{n_{ur}}.
\label{n5}
\end{eqnarray}

\noindent The other particle densities are given by $n_{ug} = n_{dr}
= n_{dg} = n_{ur}$, and $n_{sg} = n_{sr}=n_{sb}$.

The pressure and Gibbs energy per baryon of the paired deconfined
phase can also be written in terms of the same parameters:
\begin{eqnarray}
P_{(Q*_{\Delta})} &=& \sum_{fc} \frac{k^4_{fc}}{12 \pi^2}  +
\frac{\mu^4_{e}}{12 \pi^2}  + \frac{1}{\pi^2} \bar{\mu}^2
\Delta^2 - B, \\
\mu_{(Q*_{\Delta})}   &=& \sum_{fc} \frac{n_{fc} \mu_{fc}}{n_B}  +
\frac{\mu_e n_e}{n_{B}},\\
n_{B,(Q*_{\Delta})}   &=& \sum_{fc} n_{fc},
\end{eqnarray}
\noindent where $k_{fc} = (\mu^2_{fc} - m^2_{fc})^{1/2}$ and
$\bar{\mu} = \mu_{ur}$.  The chemical potentials $\mu_{fc}$ are
obtained by inverting the following set of equations,
\begin{eqnarray}
n_{fc} = \frac{\mu_{fc}^3}{3 \pi^2} + \frac{2 \Delta^2
\bar{\mu}}{\pi^2} & & f_c = u_r, u_g, d_r, d_g  \\
\mu_{fc} = (3 \pi^2 n_{fc})^{1/3} & & f_c = u_b, d_b \\
\mu_{fc} = [(3 \pi^2 n_{fc})^{2/3} + m_s^2]^{1/2} & & f_c = s_r,
s_g, s_b
\end{eqnarray}

\noindent where the number densities are given by Eqs.
(\ref{n1})-(\ref{n5}).

For sufficiently large values of $\Delta$ the energy cost invested
in forcing the 2SC pairing pattern is compensated by the gain of the
condensation energy and the preferred phase is $Q^*_{\Delta}$. For
small values of $\Delta$ the preferred state is the
$Q^*_{unp}$-phase described in Olesen \& Madsen 1994, Lugones \&
Benvenuto 1998, and Bombaci, Parenti \& Vida\~na 2004.

\section{Quantum nucleation of quark matter bubbles}

To calculate the nucleation rate of quark matter in the hadronic
medium we use the Lifshitz--Kagan quantum nucleation theory
(Lifshitz \& Kagan 1972) in the relativistic form given by Iida \&
Sato (1998). The QM droplet is supposed to be a sphere of radius
$\cal R$ and its quantum fluctuations are described by the
lagrangian
\begin{equation}
L({\cal R},{\dot {\cal R}})  = - {\cal M}({\cal R})  c^2 \sqrt{1 -
({\dot {\cal R}} /c)^2}
+ {\cal M}({\cal R}) c^2 - U({\cal R}) \, , \label{eq:eq5}
\end{equation}
\noindent where $ {\cal M}({\cal R}) $ is the effective mass of the
QM droplet, and $U({\cal R}) $ its  potential energy.

Within the Lifshitz-Kagan quantum nucleation theory, one assumes
that the phase boundary  ({\it i.e.} the droplet surface) moves
slowly compared to the high sound velocity of the medium ($\dot{\cal
R} << v_s \sim c$). Thus the number density of each phase adjust
adiabatically to the fluctuations of the droplet radius, and the
system retains pressure equilibrium between the two phases. Thus,
the droplet effective mass is given by (Lifshitz \& Kagan 1972; Iida
\& Sato 1998)

\begin{equation}
{\cal M}({\cal R}) = 4 \pi \rho_H \Big(1 - {n_{b,Q*}\over
n_{b,H}}\Big)^2 {\cal R}^3  \, , \label{eq:eq6}
\end{equation}
\noindent where $\rho_H$ is the hadronic mass density, $n_{b,H}$ and
$n_{b,Q*}$ are the baryonic number densities at a same pressure in
the hadronic and  Q*-phase,  respectively. The difference in the
Gibbs free energy is given by (Lifshitz \& Kagan 1972, Iida \& Sato
1998)

\begin{equation}
U({\cal R})   =    {4 \over 3} \pi  {{\cal R}^3} n_{b,Q*} (\mu_{Q*}
- \mu_H)   + 4 \pi \sigma {\cal R}^2   , \label{eq:eq7}
\end{equation}
\noindent where  $\mu_H$ and $\mu_{Q*}$ are the hadronic and quark
chemical potentials at a fixed pressure $P$. For comparison with
previous works notice that $\mu$ is the same as the bulk Gibbs
energy per baryon $g = (\rho + P) /n_B = ( \sum_{i} n_{i} \mu_{i} )
/n_B$. For $\sigma = 0$ we have $U({\cal R}) =$ Volume $ \times n_B
\times \Delta g$, and the bulk limit is recovered.

Notice that we neglected the term associated with the curvature
energy, and also terms connected with the electrostatic energy,
since they are known to introduce small corrections (Iida \& Sato
1998, Bombaci et. al 2004). The value of the surface tension
$\sigma$ for the interface separating the quarks and hadrons  phase
is poorly known, and typical values used in the literature range
within 10--50 MeV/fm$^2$ (Heiselberg et al. 1993, Iida \& Sato
1998). Notice that, since we are considering $\sigma$ as a
parameter, color superconductivity enters the above expressions only
trough $\mu_{Q*}$ and $n_{b,Q*}$ (see next section).


The probability of formation of a bubble having a critical radius
can be computed using a semiclassical approximation. The procedure
is rather straightforward. First one computes, using the well known
Wentzel--Kramers--Brillouin (WKB) approximation, the ground state
energy $E_0$ and the oscillation frequency $\nu_0$ of the virtual QM
drop in the potential well $U({\cal R})$. Then it is possible to
calculate in a relativistic framework the probability of tunneling
as (Iida \& Sato 1998)
\begin{equation}
p_0=\exp \Big[-{A(E_0)\over \hbar}\Big] \label{eq:eq8}
\end{equation}
\noindent where $A$ is the action under the potential barrier
\begin{equation}
A(E)  = {2\over c}\int_{{\cal R}_-}^{{\cal R}_+} \left\{ [2{\cal
M}c^2 + E -U]
             [U-E] \right\}^{1/2} d{\cal R}   \, ,
\label{eq:eq9}
\end{equation}
${\cal R}_\pm$ being  the classical turning points.

The nucleation time is then equal to
\begin{equation}
\tau = (\nu_0 p_0 N_c)^{-1}\, , \label{eq:eq10}
\end{equation}
\noindent where $N_c$ is the number of virtual centers of droplet
formation in the innermost region of the star. Following the simple
estimate given in Iida \& Sato (1998), we take  $N_c = 10^{48}$. The
uncertainty in the value of $N_c$ is expected to be within one or
two orders of magnitude. In any case, all the qualitative features
of our scenario will be not affected by the uncertainty in the value
of $N_c$.

\begin{figure}
\includegraphics[angle=0,width=9.3cm]{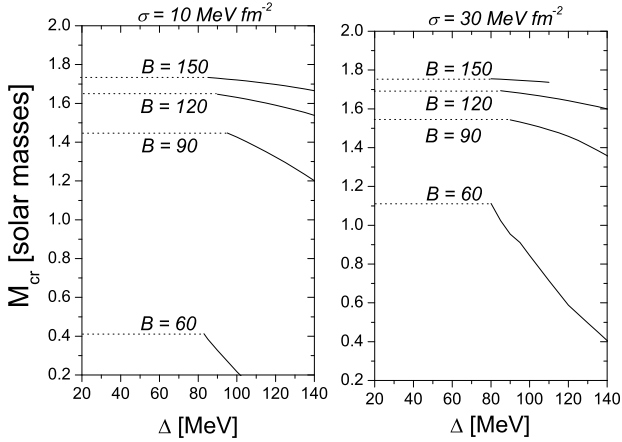}
\caption{The critical mass of hadronic stars for the nucleation of a
quark matter drop at the center. We have considered the
parametrization  GM1 (Glendenning \& Moszkowski 1991)  for the EOS
of the hadronic phase, and two different values for the surface
tension of the quark-hadron interface: $\sigma = 10$~MeV/fm$^2$ and
$\sigma = 30$~MeV/fm$^2$.  Different curves correspond to different
values of the Bag constant (indicated in units of $\mathrm{MeV
fm^{-3}}$). For small values of the pairing gap $\Delta$ the
nucleated intermediate phase is unpaired ($Q^*_{unp}$) and is
indicated with dotted line. For sufficiently large  $\Delta$ the
preferred intermediate phase is paired ($Q^*_{\Delta}$) and the
critical mass decreases substantially (solid line). The maximum mass
for hadronic stars with the GM1 EOS is $M^{max}_{HS} = 1.78
M_{\odot}$.}
\end{figure}

\begin{figure}
\includegraphics[angle=0,width=9.3cm]{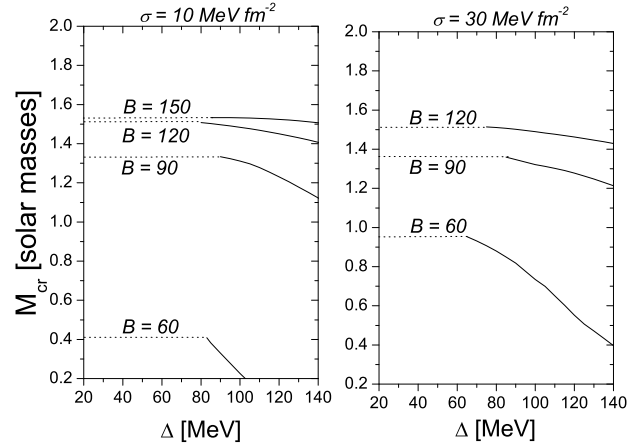}
\caption{The same as Fig. 1 but using GM3 for the hadronic phase.
The maximum mass for hadronic stars with the GM3 EOS is
$M^{max}_{HS} = 1.53 M_{\odot}$.}
\end{figure}



In order to explore the astrophysical implications of the nucleation
we shall introduce the concept of \textit{critical mass} of an
hadronic star (see e.g. Bombaci et al. 2004). In bulk, the
transition begins when the Gibbs conditions  $\Delta P = 0$ and
$\Delta \mu = 0$ are verified. However, as a consequence of the
surface effects it is necessary now to have an overpressure $\Delta
P = P - P_0 > 0$ with respect to the bulk transition point. This
overpressure will produce $\Delta \mu < 0$ in Eq. (\ref{eq:eq7})
allowing the barrier to be tunneled. Since there is an univocal
relation between the mass of a compact star and its central
pressure, we can consider that the nucleation time in Eq.
(\ref{eq:eq10}) is a function of the mass of the star. The larger
the overpressure, the easier will be to nucleate a bubble. In other
words, the larger the mass of a pure hadronic star, the shorter the
time to nucleate a quark drop at the center of the star. We define
as {\it critical mass} $M_{cr}$ of the metastable hadronic star
(HS), the value of the gravitational mass for which the nucleation
time is equal to one year: $M_{cr} \equiv M_{HS}(\tau = 1 ~{\rm
yr})$. It is worth recalling that the nucleation time given in Eq.
(\ref{eq:eq10}) is an extremely steeply function of the hadronic
star mass (see e.g. Bombaci et al. 2004). Therefore, the exact value
of the nucleation time ($\tau = 1 ~{\rm yr}$ in our case) chosen in
the definition of $M_{cr}$ is not relevant \footnote{Notice also
that, due to this tremendously steeply dependence of $\tau$ with
mass of the star, the poor knowledge of the factor $N_c$ in eq. (21)
does not modify significantly the value obtained for $M_{cr}$.}. For
example, a variation of several orders of magnitude in the choice of
$\tau$ (e.g. 1 s or $10^{17}$ s), will change $M_{cr}$ by less than
0.1 \%. Therefore, pure hadronic stars with $M_{HS}
> M_{cr}$ are very unlikely to be observed, while pure hadronic
stars with $M_{HS} < M_{cr}$ are safe with respect to a sudden
transition to quark matter. $M_{cr}$ plays the role of an {\it
effective maximum mass} for the hadronic branch of compact stars
(see the discussion in Bombaci et. al 2004). While the
Oppenheimer--Volkov maximum mass $M^{max}_{HS}$ (Oppenheimer \&
Volkov 1939)  is determined by the overall stiffness of the EOS for
hadronic matter, the value of $M_{cr}$  will depend in addition on
the properties of the intermediate non-$\beta$ stable quark phase
($Q^*_{unp}$ or $Q^*_{\Delta}$). As emphasized in LB05, the critical
mass exists even in the absence of surface effects, although its
value is smaller.

The results are shown in Figs. 1-4 where we show the critical mass
of hadronic stars as a function of the different parameters of the
equations of state. We have adopted rather common models for
describing both the hadronic and the quark phase of dense matter.
For the hadronic phase we used models which are based on a
relativistic lagrangian of hadrons interacting via the exchange of
sigma, rho and omega mesons. The parameters adopted are the standard
ones (Glendenning \& Moszkowski 1991). Hereafter we refer to this
model as the GM equation of state (EOS) \footnote{The names GM1 and
GM3 refer to the parameters given in the first and third lines of
Table 2 of Glendenning \& Moszkowski (1991). The corresponding
compressibility and coupling constants are the following. For GM1:
$K = 300 ~ \mathrm{MeV}$, $(g_{\sigma}/m_{\sigma})^2 = 11.79
~\mathrm{fm^2}$, $(g_{\omega}/m_{\omega})^2 = 7.149 ~\mathrm{fm^2}$,
$(g_{\rho}/m_{\rho})^2 = 4.411 ~\mathrm{fm^2}$, $b = 0.002947$, $c =
-0.001070$. For GM3: $K = 240 ~\mathrm{MeV}$,
$(g_{\sigma}/m_{\sigma})^2 = 9.927 ~\mathrm{fm^2}$,
$(g_{\omega}/m_{\omega})^2 = 4.820 ~\mathrm{fm^2}$,
$(g_{\rho}/m_{\rho})^2 = 4.791 ~ \mathrm{fm^2}$, $b = 0.008659$, $c
= -0.002421$. The two hadronic models include the lowest baryon
octet ($n$, $p$, $\Lambda$, $\Sigma^{+}$, $\Sigma^{0}$,
$\Sigma^{-}$, $\Xi^{-}$, and $\Xi^{0}$).}. For quark matter we used
the MIT Bag model with $m_u = m_d =0$, $m_s = 150$~MeV  and
$\alpha_s = 0$ \footnote{Other models, such as the Nambu - Jona -
Lasinio model (see Buballa 2005), can give different results,
specially in the low density regime for which the chiral symmetry is
not completely restored.}.


Before analyzing the results, it is worth to remark that there are
two qualitatively different possibilities concerning the state of
quark matter: either it is absolutely stable (i.e. it has an energy
per baryon at $P=0$ and $T=0$ smaller than the mass of the neutron)
or it is not absolutely stable. Since the actual case in nature is
unknown the so called \textit{stability windows} have been
introduced (Farhi \& Jaffe 1984). These windows show the regions in
which each possibility is realized as a function of the different
parameters of the quark model (e.g. the strange quark mass $m_s$,
the Bag constant $B$, and the pairing gap $\Delta$). The stability
windows for color-flavor locked quark matter have been presented in
Lugones \& Horvath (2002). Fixing the value of the strange quark
mass to $m_s = 150~ \mathrm{MeV}$ it is found that paired
$\beta$-stable quark matter is absolutely stable for

\begin{equation}
B <  ~73.1 ~\mathrm{MeV ~fm^{-3}} ~ \bigg[ 1 + 0.53 ~ \bigg(
\frac{\Delta}{100 ~\mathrm{MeV}} \bigg)^2 ~  \bigg]  .
\end{equation}

\noindent The above expression is an approximation to order $m_s^2$,
accurate within a few percent in the range of interest (see Eq. (17)
of Lugones \& Horvath 2002). If quark matter is not absolutely
stable, the stars containing quark matter are \textit{hybrid stars},
i.e. containing $\beta$-stable quark matter only at their interiors.
If quark matter is absolutely stable, all the stars containing quark
matter are \textit{strange stars}, i.e. made up of $\beta$-stable
quark matter from the center to the surface.



\begin{figure}
\includegraphics[angle=0,width=9.3cm]{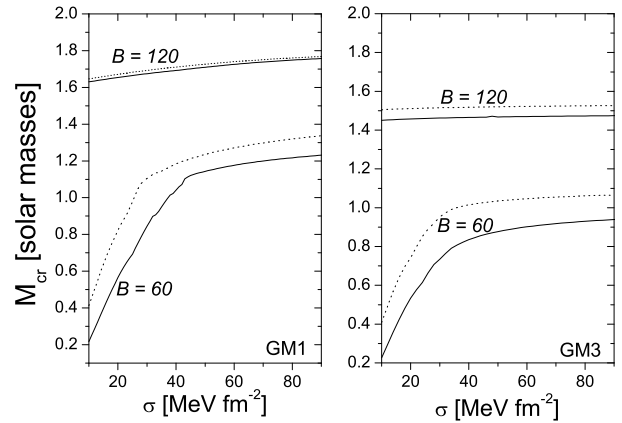}
\caption{The critical mass of hadronic stars as a function of the
surface tension $\sigma$ of the quark-hadron interphase. We show the
results for two values of the Bag constant $B$ (indicated in units
of $\mathrm{MeV ~fm^{-3}}$). Dotted lines correspond to unpaired
quark matter, solid lines to paired quark matter with $\Delta = 100~
\mathrm{MeV}$. }
\end{figure}

In Figs. 1 and 2 the effect of the pairing gap $\Delta$ is seen
clearly. For small values of $\Delta$ the nucleated intermediate
phase is in an unpaired state (dotted line). For sufficiently large
values of $\Delta$ the gain of the condensation energy favors the
nucleation of a paired state ($Q^*_{\Delta}$)  and the critical mass
decreases substantially due to the lower energy cost of the phase.
Notice that the effect of pairing is stronger for small values of
the Bag constant $B$, where it can change $M_ {cr}$ by more than a
factor of 2. On the other hand, the variation of the critical mass
due only to pairing effects (i.e. at constant $B$) is smaller than a
$\sim 20 \%$ in the regime of large $B$ that corresponds to hybrid
stars (i.e. non absolute stability of quark matter).


In figure 3 we see the effect of the surface tension $\sigma$ on the
critical mass $M_ {cr}$. The effect of $\sigma$ is small for $B$
larger than $\sim ~ 100 ~ \mathrm{MeV ~ fm^{-3}}$. Therefore, in
this regime of $B$ the bulk limit is a good approximation (c.f.
LB05). For $B$ smaller than $\sim ~ 100 \mathrm{MeV ~ fm^{-3}}$ the
effects of $\sigma$ are important, and increase their relevance as
$B$ decreases. This can be seen in Fig. 3 in the case for $B = 60 ~
\mathrm{MeV ~ fm^{-3}}$, where $M_ {cr}$ changes by a factor of 4 in
the most plausible range for $\sigma$ (between 10 and 30
$\mathrm{MeV ~ fm^{-2}}$). However, notice that the curves tend to
``saturate'' for larger  $\sigma$.


\begin{figure}
\includegraphics[angle=0,width=9.3cm]{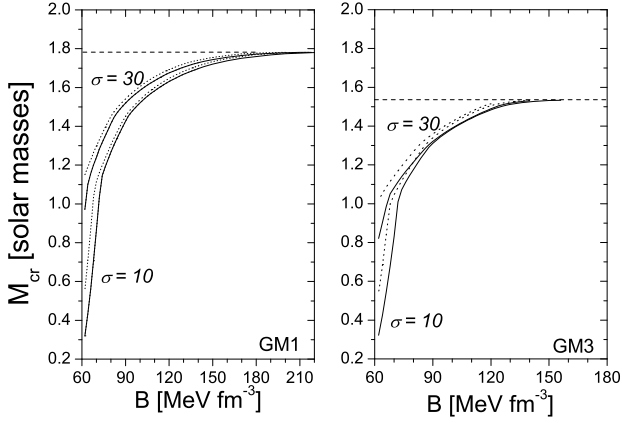}
\caption{The critical mass of hadronic stars as a function of the
Bag constant $B$.  The results are shown for $\sigma =$ 10 and 30
$\mathrm{MeV ~fm^{-2}}$. Dotted lines correspond to unpaired quark
matter, solid lines to paired quark matter with $\Delta = 100~
\mathrm{MeV}$. The  dashed horizontal line indicates the maximum
mass of hadronic stars for each hadronic equation of state.}
\end{figure}

In Fig. 4 we explore the dependence of the results on the Bag
constant $B$. We can  identify three qualitatively different
regions:

(1) \textit{Region of Large $B$:} For $B$ larger than a critical
value ($\sim 200 ~\mathrm{MeV ~ fm^{-3}}$  for GM1 and $\sim 150
~\mathrm{MeV ~ fm^{-3}}$ for GM3)  the critical mass $M_{cr}$ is no
longer smaller than the maximum mass for hadronic stars
$M^{max}_{HS}$. This implies that quark stars cannot form by means
of quantum nucleation in this region of the parameter space.

(2) \textit{Region of intermediate $B$ corresponding to hybrid
stars:} For $B$ between $\sim 100 ~\mathrm{MeV ~ fm^{-3}}$ and $\sim
150-200 ~\mathrm{MeV ~ fm^{-3}}$ the critical mass $M_{cr}$ is very
close (but smaller than) the maximum mass of hadronic stars. Notice
that for this range of $B$, stars containing quark matter are hybrid
since quark matter in $\beta$-equilibrium has an energy per baryon
(at $P=0$) smaller than the neutron mass (we assume for simplicity
that $\beta$-stable quark matter is in a color flavor locked state).

(3) \textit{Region of  $B$ corresponding to strange stars:} For
sufficiently small $B$ quark matter in $\beta$-equilibrium is
absolutely stable (see Eq. (22)).  As apparent from Fig. 4, the
critical mass $M_{cr}$ is strongly dependent on $B$ in this case.

\section{Energy released in the conversion}

The amount of energy that can be released in the conversion of a
pure hadronic star (HS) into a strange star or an hybrid star has
been calculated in several previous works (Bombaci \& Datta 2000,
Berezhiani et al. 2003, Bombaci et al. 2004, Drago, Lavagno \&
Pagliara 2004). Moreover, Drago, Lavagno \& Pagliara (2004)
calculated the available energy including the effect of color
superconductivity in the final star. However, since the critical
masses in that work were not calculated employing the here studied
intermediate non-$\beta$-stable phase, we present here new
self-consistent results.

As in previous works, we define $\Delta E$ as the difference between
the gravitational mass of the initial hadronic star and that of the
final (hybrid or strange) star having the same baryonic mass:

\begin{equation}
\Delta E = [M_G(HS) -  M_G(QS)]
c^2. \end{equation}

\noindent According to the definition,  $\Delta E$ is the total
energy that can be released in the conversion. This energy can be
liberated in an explosive manner (and give raise to GRBs) or in a
less violent way (in which case observable signals would be less
spectacular), but this depends on complex processes that are out of
the scope of the present work (see e.g. Lugones, Ghezzi, de Gouveia
Dal Pino \& Horvath 2002,  Ouyed, Rapp and Vogt 2005, Paczynski and
Haensel 2005, Bhattacharyya, Ghosh and Raha 2005,  and references
therein).

The results are shown in Fig. 5 and 6 where we show $\Delta E$ as a
function of the paring gap $\Delta$ for different values of the Bag
Constant $B$, using the equations of state GM1 and GM3 for hadronic
matter, and assuming that the $\beta$-stable quark phase (for the
final stellar configuration) is in the color flavor locked (CFL)
state. We assume that a given hadronic star can convert into a quark
star provided it has the critical mass. Those with a smaller mass
cannot convert, and those with larger mass are already converted.
Each point of the curves of Figs. 5 and 6 indicate the energy
released in the conversion of an hadronic star having the critical
mass corresponding to that $B$ and $\Delta$. This hadronic star is
converted into a quark star (having the same barionic mass) that can
be a strange star or a hybrid star depending also on the values of
$B$ and $\Delta$ (this can be known with the help of Eq. 22).

The released energy is mainly the result of two competing effects.
For $\Delta$ above a threshold value (see Figs. 1 and 2), pairing
tends to diminish $M_{cr}$, and therefore, decreases the released
energy since the conversion happens at the critical mass. On the
other hand, there is an effect on the gravitational mass of the
hybrid or strange star due to the condensation term in the equation
of state of CFL quark matter. Depending on $B$ and $\Delta$ the
equation of state can be stiffer or softer than the unpaired case
(see e.g. Lugones and Horvath 2003) and therefore $\Delta E$ can be
either an increasing or a decreasing function of $\Delta$.

The dotted part of the curves in Figs. 5 and 6 corresponds to the
case in which the intermediate non-$\beta$-stable quark phase is
unpaired. Therefore, the critical mass is independent of $\Delta$
and the variations in $\Delta E$ are due only to the variation in
the stiffness of the equations of state, that changes the
gravitational mass of the final star. The full-line part of the
curves corresponds to the case in which the intermediate phase is
paired. Since the critical mass is a decreasing function of $\Delta$
(see Figs. 1 and 2) the released energy tends to decrease, specially
for low values of $B$, for which the decrease of $M_{cr}$ is
stronger (see e.g. the case of $B = 60 ~ \mathrm{MeV fm^{-3}}$). For
larger $B$ the decrease in $M_{cr}$ is not enough to compensate the
effect of the stiffness of the equations of state. In fact, for very
large $\Delta$, $\Delta E$ is always an increasing function of
$\Delta$ because the gain in the condensation energy compensates all
other effects.

\begin{figure}
\includegraphics[angle=0,width=8.5cm]{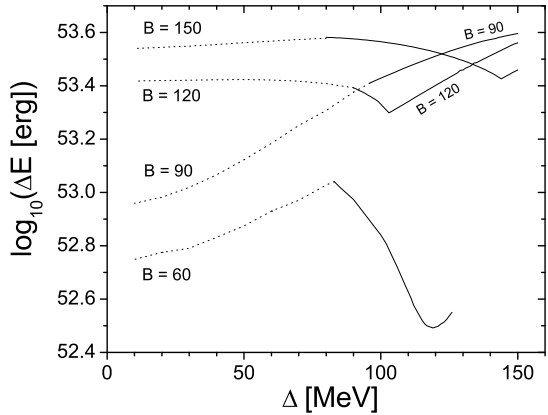}
\caption{The total energy released in the conversion of an hadronic
star (having the critical mass) into a quark star (hybrid star or
strange star depending on the values of $B$ and $\Delta$). In dotted
lines we show the cases in which the intermediate non-$\beta$-stable
quark phase is unpaired, while the full lines corresponds to the
case in which the intermediate phase is paired (see more details in
the text). In this figure we used the GM1 EOS for hadronic matter.
The critical mass is calculated with $\sigma = 10 ~ \mathrm{MeV
fm^{-2}}$. The Bag constant is indicated in units of $\mathrm{MeV
fm^{-3}}$.}
\end{figure}

\begin{figure}
\includegraphics[angle=0,width=8.5cm]{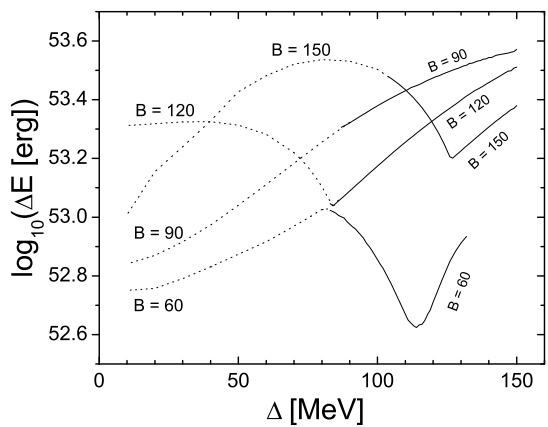}
\caption{The same as Fig. 5 but employing the GM3 EOS for hadronic
matter.}
\end{figure}

\section{Discussion and Conclusions}

In the present work we have studied the nucleation of quark matter
drops at the center of cold deleptonized neutron stars. These drops
can be made up by unpaired quark matter or by color superconducting
quark matter. The nature of the nucleated phase depends on the
details of the equations of state for quark and hadronic matter, and
is relevant in the determination of the critical mass $M_{cr}$ of
the hadronic stars above which a transition to quark matter is
possible. We have calculated $M_{cr}$  as a function of the several
parameters of the quark model, and for different parametrization of
the hadronic equations of state.  In general, the dependence of
$M_{cr}$ on  $B$, $\Delta$ and $\sigma$ is mild if the parameters of
the quark model correspond to hybrid stars, and strong if they
correspond to strange stars. Also, the critical mass always
decreases with $\Delta$, and increases with $B$ and $\sigma$.

As stated before, if quark matter is not absolutely stable, the
stars containing quark matter are hybrid stars, i.e. containing
$\beta$-stable quark matter only at their interiors. On the other
hand,  if quark matter is absolutely stable, all the stars
containing quark matter are strange stars, i.e. made up of
$\beta$-stable quark matter from the center up to the surface.
Moreover, it has been argued that if $\beta$-stable quark matter is
absolutely stable, then \textit{all neutron stars} should be made up
by quarks. The reason is that cosmic ray strangelets will be created
after the merging of a strange star with another compact star in a
binary system. The rate of binary mergers in the galaxy appears to
be large enough so that even a conservative estimate gives a large
galactic production rate of strangelets ( $10^{-10} M_{\odot}/yr$),
and correspondingly, a large flux in the interstellar medium. When a
strangelet penetrates a neutron star, it will grow by absorbing free
neutrons, converting the whole star into strange quark matter.
Similarly, all massive stars that accumulate a strangelet in its
core will give birth to a strange star rather than a neutron star
after the explosion as a core collapse supernova. Because of these
arguments, it has been argued that if some neutron stars are
undoubtedly identified as being conventional, the inevitable
conclusion is that strange quark matter is not absolutely stable and
there exist only conventional neutron stars or (eventually) hybrid
stars (see Caldwell \& Friedman 1991, Balberg 2005). However, notice
that it may happen that the strangelet contamination of the galaxy
is suppressed, for example, if strange quark stars are disfavored in
binaries (Belczynski et al. 2002 and references therein). On the
other hand, strangelets could disintegrate when impacting onto the
external layers of the star. But, the physics of strangelet
fragmentation is largely unknown and it is not clear whether it is
effective for all incident energies (specially for those strangelets
impacting with low energy). In conclusion, if quark matter is in
fact absolutely stable, it is not clear whether the conversion would
be triggered from inside (quantum nucleation) or from outside
(strangelet contamination), and therefore, it is uncertain whether
the critical mass studied in this paper is meaningful in the
corresponding region of the parameter space ($B$, $\Delta$, etc.).

On the other hand, if quark matter is not absolutely stable,
strangelets do not exist. Therefore, the trigger of the conversion
to quark matter cannot come from outside the neutron star, and the
here studied mechanism is the most plausible one to trigger the
conversion in cold deleleptonized neutron stars. In this context, a
relevant result presented here is that for a large region of the
parameter space \textit{corresponding to hybrid stars}, the critical
mass is very near (but smaller than) the maximum mass of hadronic
stars. This means that all compact stars with masses up to $M_{cr}$
should be normal hadronic stars while those with masses above
$M_{cr}$ may be hybrid stars (if they do not collapse into a black
hole). In the rest of the parameter space corresponding to hybrid
stars the critical mass $M_{cr}$ is no longer smaller than the
maximum mass for hadronic stars $M^{max}_{HS}$. This implies that
quark stars cannot form by means of quantum nucleation in this
region of the parameter space, and probably they don't exist at all
(if other mechanism of quark matter formation does not operate).

We have also calculated the amount of energy that can be released in
the conversion of an hadronic star into a quark or hybrid star
(assuming that the $\beta$-stable quark matter phase is in a color
flavor locked state). The total released energy is in the range $3
\times 10^{52} \mathrm{erg}$ - $4 \times 10^{53} \mathrm{erg}$, and
therefore it is sufficient to power a gamma ray burst.

\end{document}